\documentclass[preprint,aps,showkeys,preprintnumbers,amsmath,amssymb,footinbib]{revtex4-2}
\usepackage{mathrsfs}
\usepackage{amssymb}
\usepackage{graphicx}
\usepackage{hyperref}
\usepackage{ulem}
\usepackage{tabularx}
\usepackage{array}
\usepackage{color}

\begin{document}

\title{Probability Density in Relativistic Quantum Mechanics}

\author{Taeseung Choi}
 \email{tschoi@swu.ac.kr}
\affiliation{Division of General Education, Seoul Women's University, Seoul 139-774, Korea}
\affiliation{School of Computational Sciences, Korea Institute for Advanced Study, Seoul 130-012, Korea}

\author{Yeong Deok Han}
\email{ydhan2@hanmail.net}
\affiliation{Department of Computer Science and Engineering, Woosuk University, Jincheon, Chungbuk, 27841, Korea}

\begin{abstract}

In the realm of relativistic quantum mechanics, we address a fundamental question: Which one, between the Dirac or the Foldy-Wouthuysen density, accurately provide a probability density for finding a massive particle with spin $1/2$ at a certain position and time. Recently, concerns about the Dirac density's validity have arisen due to the Zitterbewegung phenomenon, characterized by a peculiar fast-oscillating solution of the coordinate operator that disrupts the classical relation among velocity, momentum, and energy. To explore this, we applied Newton and Wigner's method to define proper position operators and their eigenstates in both representations, identifying 'localized states' orthogonal to their spatially displaced counterparts. Our analysis shows that both densities could represent the probability of locating a particle within a few Compton wavelengths. However, a critical analysis of Lorentz transformation properties reveals that only the Dirac density meets all essential physical criteria for a relativistic probability density. These criteria include covariance of the position eigenstate, adherence to a continuity equation, and Lorentz invariance of the probability of finding a particle. Our results provide a clear and consistent interpretation of the probability density for a massive spin-$1/2$ particle in relativistic quantum mechanics.

\end{abstract}


\maketitle

 \section{ Introduction}
Relativistic quantum mechanics (QM) has profoundly advanced our understanding of the quantum behavior of particles at high velocities, with the Dirac theory playing a crucial role in describing massive spin $1/2$ fermions, such as electrons \cite{Dirac1928,PQMDirac}. Central to this theory is the Dirac density $\psi^\dagger_D(x)\psi_D(x)$, derived from the Dirac spinor field $\psi_D(x)$ in the Dirac representation, where $x$ represents 4-position vector $x^\mu$. This density is crucial as it represents the probability of finding an electron per unit volume at a specific position at time $t=x^0$ (with $c=1$).  Its widespread acceptance stems from its fulfillment of key properties required for a probability density, namely positiveness and being the $0$-th component of a 4-vector that satisfies the continuity equation \cite{PQMDirac}. Consequently, the Dirac density has found extensive application as a probability density across diverse domains, ranging from vortex physics \cite{Bliokh2017,Lloyd2017} to quantum chemistry \cite{Reiher2009,Mastalerz2008}. 

However, an alternative argument has been proposed \cite{SilenkoPRL18, SilenkoPRA20}, suggesting that the Dirac density is incorrect and advocating for the use of the density in the Foldy-Wouthuysen (FW) representation \cite{FW50}, denoted by $\psi^\dagger_{FW}(x)\psi_{FW}(x)$. This proposition is based on the ease of associating position and spin operators with their classical counterparts in the FW representation \cite{SilenkoPRA20}. Recent debates between Birula \textit{et al.} and Silenko \textit{et al.} have focused on determining which density, either $\psi_D^\dagger(x)\psi_D(x)$ or $\psi_{FW}^\dagger(x) \psi_{FW}(x)$, more accurately describes the probability of locating a particle per unit volume at a specific position ${\bf x}$ at time $x^0$ \cite{BirulaComment,SilenkoReply}. Birula \textit{et al.} argue that the FW density $\psi_{FW}^\dagger(x) \psi_{FW}(x)$ fails to satisfy the continuity equation and does not accurately reproduce the beam shape of a relativistic vortex. Conversely, Silenko \textit{et al.} contend that the Dirac representation introduces distortions in the classical relationship among velocity, momentum, and energy, resulting in the unusual rapid oscillations of the Dirac position operator $\hat{\textbf x}$, which is the operator representative of the spatial coordinate in the Dirac representation---a phenomena known as the Zitterbewegung \cite{Schrodinger30}.

A fundamental expectation in QM is that a position operator should reproduce the classical relationship between velocity, momentum, and energy, as reflected in the expectation values defined by a probability density. However, the occurrence of Zitterbewegung in the Dirac position operator in the Dirac representation suggests that the Dirac position operator may not be ideal. This observation does not imply that the Dirac representation is invalid for representing probability as a function of position. Rather, it underscores the need for finding a position operator in the Dirac representation that accurately reproduces the classical relationships among velocity, momentum, and energy.  

The interpretation of the probability density to find a particle at a specific position in relativistic QM necessitates the use of a proper position operator. In this paper, we begin by identifying potential candidates for position eigenstates in both the Dirac and FW representations. Our investigation involves examining sets of localized states that conform to the locality criteria established by Newton and Wigner \cite{NW49}. The localized states naturally introduce proper position operators in both representations by being position eigenstates. We demonstrate that, upon applying the FW transformation to the proper position operator in the Dirac representation, it becomes equivalent to the proper position operator in the FW representation, specifically when this operator acts on the positive-energy (particle) subspace. This finding suggests that the choice of the proper position operator does not inherently favor one representation over the other for generating a suitable probability density. Building on this understanding, we delve into the physical requirements for the probability density associated with the two candidates: $\psi_D^\dagger(x)\psi_D(x)$ and $\psi_{FW}^\dagger(x) \psi_{FW}(x)$. Our focus is on evaluating their adherence to the continuity equation and the Lorentz invariance of the probability of finding a particle.


The paper is organized as follows. In Sec. \ref{sec:PEG} we obtain the position eigenstates in the Dirac and the FW representation, localized at a specific position under the Newton and Wigner (NW) locality condition \cite{NW49}. We explore the covariance of the position eigenstates in the two representations under Lorentz transformations. In Sec. \ref{sec:RPP} we study the physical requirements for the probability density: the continuity equation and the Lorentz invariance of the probability to find a particle. We summarize our results in Sec. \ref{sec:Conc}.

\section{Position eigenstates in the Dirac and the FW representations}
\label{sec:PEG}

To interpret $\psi^\dagger(x)\psi(x)$, with $x=(x^0, {\bf x})$, as the probability of finding a particle at a specific position ${\textbf x}$ at time $t=x^0$ per unit volume, it is essential in both non-relativistic and relativistic QM to properly define a position operator and its corresponding localized eigenstate. In the Dirac representation, the coordinate vector operator $\hat{\textbf x}$, referred to as the Dirac position operator, exhibits peculiar behavior in the context of a free particle's position, primarily attributed to the phenomenon of Zitterbewegung. This behavior, which complicates the interpretation of $\hat{\textbf x}$ as a proper position operator, has been notably discussed by Silenko \textit{et al.} \cite{SilenkoReply}. In contrast, the coordinate vector operator in the FW representation shows no Zitterbewegung and aligns well with the classical relativistic mechanics’ relationship among velocity, momentum, and energy. Silenko \textit{et al.} argued that this alignment supports the FW density in the FW representation as the appropriate probability density in relativistic QM.

While the presence of Zitterbewegung raises concerns about the suitability of the Dirac position operator $\hat{\textbf x}$ as a proper position operator, this issue does not necessarily reflect a fundamental inadequacy of the Dirac density itself as a proper probability density. Our aim in this section is to explore suitable position operators that conform to Newton-Wigner's locality criteria \cite{NW49} within both the Dirac and FW representations. In the following paragraphs, we begin by reviewing our recent study \cite{Our21PO, Our21SP}, where we successfully identified and characterized a position operator with localized eigenstates in the Dirac representation. 


For the sake of completeness and to elucidate our methodology, we briefly review how our position eigenstates and operators in the Dirac representation are derived, aligning with the NW locality criteria. In relativistic QM, the momentum representation is commonly used as it categorizes states by both momentum and spin. Using this representation, we constructed localized position eigenstates such as $e^{ip \cdot x}\psi_{+\epsilon}(p,\lambda)$ and $e^{-ip\cdot x}\psi_{-\epsilon}(p,\lambda)$ for both particles and antiparticles within the Dirac framework \cite{Our21PO}. These states are characterized by momentum $p=p^\mu=(p^0,{\textbf p})$ and spin $\lambda=\pm 1/2$. Here, $p\cdot x=p^\mu x_\mu$ denotes the scalar product in four-dimensional spacetime, using the Einstein summation convention. These eigenstates share the same eigenvalue ${\textbf x}$ at the time slice $t=x^0$ (hypersurface of simultaneity). 

The spinors $\psi_{+ \epsilon}(p,\lambda)$ and $\psi_{- \epsilon}(p,\lambda)$ represents positive-energy solutions for the particle Hamiltonian $H_{P}= \boldsymbol{\alpha}\cdot{\textbf p}+\beta m$ and the antiparticle Hamiltonian $H_{AP}=\boldsymbol{\alpha}\cdot{\textbf p}-\beta m$ in the Dirac formalism, respectively. Here, $\boldsymbol{\alpha}\cdot{\textbf p}=\alpha^k p^k$ (for $k=1,2,3$) and the Dirac matrices are defined as $\alpha^k=\left( \begin{array}{cc} 0 & \sigma^k \\ \sigma^k & 0 \end{array}\right)$ and $\beta= \left( \begin{array}{cc} 1 & 0 \\ 0 & -1 \end{array}\right)$ in the standard Dirac representation, with $\sigma^k$ being the Pauli matrices \cite{PQMDirac}. These spinors satisfy the eigenvalue equations:
\begin{subequations}
\begin{eqnarray}
H_{P}\, \psi_{+\epsilon}(p,\lambda) = E_{\textbf p} \, \psi_{+\epsilon}(p,\lambda) \\
H_{AP}\, \psi_{-\epsilon}(p,\lambda) = E_{\textbf p}\, \psi_{-\epsilon}(p,\lambda),
\end{eqnarray}
\end{subequations}
where $E_{\textbf p}=\sqrt{ {\textbf p}^2+m^2}$ represents the relativistic energy with ${\textbf p}^2={\textbf p}\cdot{\textbf p}$.  

To demonstrate that the states $e^{ip \cdot x}\psi_{+\epsilon}(p,\lambda)$ and $e^{-ip\cdot x}\psi_{-\epsilon}(p,\lambda)$ are localized, we apply the NW locality criteria \cite{NW49}. According to this criterion, these states must be orthogonal to any states that are spatially translated within the same time slice. The orthogonality relations are defined through the following scalar products:
\begin{subequations}
\begin{eqnarray}
\label{eq:SCP}
\int dp \, \psi_{+\epsilon}^\dagger(p,\lambda) e^{-i{\textbf p}\cdot({\textbf x}-{\textbf x'})} \psi_{+\epsilon}(p,\lambda) &=& \delta({\textbf x}-{\textbf x'}), \\ 
\int dp \, \psi_{-\epsilon}^\dagger(p,\lambda) e^{i{\textbf p}\cdot({\textbf x}-{\textbf x'})} \psi_{-\epsilon}(p,\lambda) &=& \delta({\textbf x}-{\textbf x'}),
\end{eqnarray}
\end{subequations}
demonstrating that these states are orthogonal to states located at a different position ${\textbf x'}$. These orthogonality relations are derived by using the fact that $\psi_{+\epsilon}^\dagger(p,\lambda) \psi_{+\epsilon}(p,\lambda')=\psi_{-\epsilon}^\dagger(p,\lambda) \psi_{-\epsilon}(p,\lambda')=(E_{\textbf p}/m)\, \delta_{\lambda, \lambda'}$, where, $dp =  \frac{d^3 {\textbf p}}{(2\pi)^3} \frac{m}{E_{\textbf p}}$ represents the Lorentz invariant integral measure on the mass-shell, with $d^3 {\textbf p}=dp^1 dp^2 dp^3$.

The localized states $e^{ip \cdot x}\psi_{+\epsilon}(p,\lambda)$ and $e^{-ip\cdot x}\psi_{-\epsilon}(p,\lambda)$ for a particle and an antiparticle respectively, naturally lead to the formulation of the particle position operator $X^k_P$ and the antiparticle position operator $X^k_{AP}$ in the momentum representation of the Dirac representation, as detailed in our work \cite{Our21PO}:
\begin{subequations}
\begin{eqnarray}
\label{eq:PPO}
X^k_{P}&=& M(L_{\textbf p}) (i\partial_{p^k}) M^{-1}(L_{\textbf p}) \\
\label{eq:APPO}
X^k_{AP}&=& M(L_{\textbf p}) (-i\partial_{p^k}) M^{-1}(L_{\textbf p}),
\end{eqnarray}
\end{subequations}
where $M(L_{\textbf p})$ is the spinor representation of the standard Lorentz boost $L_{\textbf p}$ that transforms the momentum as $p^\mu =(L_{\textbf p})^\mu_{\phantom{\mu}\nu}k^\nu$ with $k^\nu=(m,{\textbf 0})$. 

These position operators are designed to satisfy the eigenvalue equations:  
\begin{subequations}
\begin{eqnarray}
\label{eq:PEVE}
X^k_P \, e^{ip\cdot x} \psi_{+\epsilon}(p,\lambda) &=& x^k e^{ip\cdot x} \psi_{+\epsilon}(p,\lambda)\\
\label{eq:APEVE}
X^k_{AP} \, e^{-ip\cdot x} \psi_{-\epsilon}(p,\lambda) &=& x^k e^{-ip\cdot x} \psi_{-\epsilon}(p,\lambda).
\end{eqnarray}
\end{subequations}
These eigenvalue equations can be readily verified by utilizing the relation:
\begin{eqnarray}
\psi_{\pm \epsilon}(p,\lambda) = M(L_{\textbf p}) \psi_{\pm \epsilon}(k,\lambda),
\end{eqnarray} 
which connects the spinors in the frame of momentum ${\textbf p}$ and the rest frame with momentum ${\textbf k}={\textbf 0}$. This relation shows how the spinors transform under boosts. 

It is important to note that $X^k_P$ and $X^k_{AP}$ in Eqs. (\ref{eq:PPO}) and (\ref{eq:APPO}) differ from the Dirac position operator $\hat{x}^k$, which is represented as $i\partial_{p^k}$ for a particle and $-i\partial_{p^k}$ for an antiparticle in the momentum representation of the Dirac formalism. Our study \cite{Our21PO} has demonstrated that $X^k_P$ and $X^k_{AP}$ do not exhibit Zitterbewegung and accurately reproduce the classical relationship among velocity, momentum, and energy. This property is expected for a proper position operator, unlike the Dirac position operator, which does not satisfy these classical relations due to the Zitterbewegung phenomenon.

In order to qualify as a proper relativistic position operator, its eigenstates must exhibit form invariance (covariance) across all reference frames. The covariance of the particle and the antiparticle position eigenstates in the Dirac representation under a general Lorentz transformation $\Lambda$ has been demonstrated in our previous work by the following Lorentz transformations \cite{Our21PO}:
\begin{subequations}
\begin{eqnarray}
\label{eq:LTPE}
&&e^{i{ p}\cdot{ x}}\psi_{+\epsilon}(p,\lambda) \\ \nonumber 
&\phantom{\longrightarrow}& \phantom{\longrightarrow} \longrightarrow e^{i{  p}\cdot {\Lambda^{-1} x'}}M(\Lambda)\psi_{+\epsilon}(p,\lambda)= e^{i{\Lambda  p}\cdot { x'}}\sum_{\lambda'} \mathcal{D}_{\lambda' \lambda}(R) \psi_{+\epsilon}(\Lambda p, \lambda'), \\ 
\label{eq:LTAPE}
&&e^{-i{ p}\cdot{ x}}\psi_{-\epsilon}(p,\lambda) \\ \nonumber &\phantom{\longrightarrow}& \phantom{\longrightarrow} \longrightarrow  e^{-i{  p}\cdot   {\Lambda^{-1} x'}}M(\Lambda)\psi_{-\epsilon}(p,\lambda)=e^{-i{\Lambda  p}\cdot { x'}} \sum_{\lambda'} \mathcal{D}_{\lambda' \lambda}(R) \psi_{-\epsilon}(\Lambda p, \lambda').
\end{eqnarray}  
\end{subequations}
Here, $\Lambda p=\Lambda^\mu_{\phantom{\mu}\nu}p^\nu$ and $x'=\Lambda x$ represent the transformed momentum and position, respectively, under the Lorentz transformation $\Lambda$. The 4-dimensional matrix $\mathcal{D}_{\lambda' \lambda}(R)$ denotes the spinor representation of the Wigner rotation $R=L^{-1}(\Lambda p) \Lambda L(p)$, and $M(\Lambda)$ signifies the spinor representation of the general Lorentz transformation $\Lambda$ \cite{Our21PO,Our21SP}.

The resultant spinors $ \sum_{\lambda'} \mathcal{D}_{\lambda' \lambda}(R) \psi_{\pm\epsilon}(\Lambda p, \lambda')$ in Eqs. (\ref{eq:LTPE}) and (\ref{eq:LTAPE}) represent the eigen-spinors for the particle and antiparticle Hamiltonians in the transformed frame of momentum $\Lambda {\textbf p}$. The direction of spin can change from the original spin, as indicated by the spinor representation of Wigner rotation $\mathcal{D}_{\lambda' \lambda}(R)$. Consequently, the transformed position eigenstates in Eqs. (\ref{eq:LTPE}) and (\ref{eq:LTAPE}) remain localized in the transformed reference frame of momentum $\Lambda {\textbf p}$ after undergoing a Lorentz transformation $\Lambda$ and become the position eigenstates in the new reference frame with new eigenvalue ${\bf x'}$. The change of the direction of spin does not affect their localized nature and the fact that the transformed state becomes the position eigenstate. Hence, the Lorentz transformations of the position eigenstates in Eqs. (\ref{eq:LTPE}) and (\ref{eq:LTAPE}) imply the form invariance (covariance) of the position eigenstates under Lorentz transformations across different frames. 


The eigenvalue equations for the position operators in Eqs. (\ref{eq:PEVE}) and (\ref{eq:APEVE}) reveal that the transformed position eigenstates acquire the new position eigenvalue ${x'}^k=\Lambda^k_{\phantom{k}\mu}x^\mu$ for both the particle and antiparticle in the transformed frame. This new eigenvalue represents the spatial coordinate, covariantly transformed under the Lorentz transformation $\Lambda$. This observation supports the covariance of the particle position operator $X^k_P$ and the antiparticle position operator $X^k_{AP}$, confirming their suitability as proper position operators within the framework of relativistic QM.  

Now, let us delve into a localized state and its associated position operator within the FW representation. The FW representation is characterized by FW particle and antiparticle spinors $u_{\pm \epsilon}(p,\lambda)$, each distinguished by having only upper and lower components. These spinors are obtained by applying the FW transformation operator $U_{FW}({\textbf P})$ on the Dirac particle and antiparticle spinors $\psi_{\pm \epsilon}(p,\lambda)$. Importantly, the action of the momentum operator ${\textbf P}$ in the FW transformation $U_{FW}({\textbf P})$ on a momentum eigenstate yields opposite eigenvalues ${\textbf p}$ and $-{\textbf p}$ for particles and antiparticles, respectively, like in the Dirac representation. As a result, 
\begin{subequations}
\begin{eqnarray}
\label{eq:FWP}
u_{+\epsilon}(p,\lambda) = U_{FW}({\textbf p}) \psi_{+\epsilon}(p,\lambda) \\
\label{eq:FWAP}
u_{-\epsilon}(p,\lambda) = U^\dagger_{FW}({\textbf p}) \psi_{-\epsilon}(p,\lambda),
\end{eqnarray} 
\end{subequations}
where $U_{FW}({\textbf p})$ and its adjoint $U^\dagger_{FW}({\textbf p})$ are defined as \cite{FW50}: 
\begin{eqnarray}
U_{FW}({\textbf p})=e^{-\gamma^0 \gamma^5 \boldsymbol{\Sigma}\cdot{\boldsymbol{\xi}}/2} \mbox{ and } U^\dagger_{FW}({\textbf p})=e^{\gamma^0 \gamma^5 \boldsymbol{\Sigma}\cdot{\boldsymbol{\xi}}/2}.
\end{eqnarray}
Given that $U_{FW}({\textbf p})$ is unitary operator, it follows that $U^\dagger_{FW}({\textbf p})=U^{-1}_{FW}({\textbf p})=U_{FW}(-{\textbf p})$.
It is noteworthy that the capital letter ${\textbf P}$ denotes an operator, while the lowercase ${\textbf p}$ denotes a variable.

The FW particle and antiparticle spinors, $u_{\pm\epsilon}(p,\lambda)$, can be interpreted as states localized at the origin at time $x^0=0$. This interpretation is supported by their satisfaction of the NW locality condition:
\begin{subequations}
\begin{eqnarray}
\label{eq:NWFWP}
\int dp \,u_{+\epsilon}^\dagger(p,\lambda) e^{-i{\textbf p}\cdot{\textbf x}} u_{+\epsilon}(p,\lambda) &=&\delta({\textbf x}), \\
\label{eq:NWFWAP}
\int dp \,u_{-\epsilon}^\dagger(p,\lambda) e^{i{\textbf p}\cdot{\textbf x}} u_{-\epsilon}(p,\lambda) &=&\delta({\textbf x}),
\end{eqnarray}
\end{subequations}
utilizing the fact that $u^\dagger_{\pm \epsilon}(p,\lambda)u_{\pm \epsilon}(p,\lambda)=E_{\textbf p}/m$ parallels that of $\psi^\dagger_{\pm \epsilon}(p,\lambda) \psi_{\pm \epsilon}(p,\lambda)$, due to the unitarity of the FW transformation operator $U_{FW}({\textbf p})$. Consequently, the localized states at a position ${\textbf x}$ at time $x^0$ are represented by:
\begin{eqnarray}
\label{eq:FWPOEGP}
e^{i{ p}\cdot{ x}} u_{+\epsilon}(p,\lambda)  \mbox{ and }
e^{-i{ p}\cdot{ x}} u_{-\epsilon}(p,\lambda)
\end{eqnarray}
for the particle and the antiparticle FW spinor spaces, respectively. 

In our subsequent analysis, we will primarily concentrate on the particle case. Within the framework of relativistic QM, the number of particles and antiparticles is fixed, typically involving a single particle. A key aspect to note is that in both the FW and Dirac representations, the subspaces corresponding to particles and antiparticles remain distinct and unchanged after any physical operation \cite{Our21PO, Our21SP}. This means that all fundamental physical observables, such as position, momentum, and spin, maintain the separation between particle and antiparticle characteristics. This distinction allows us to focus our analysis on particles with the understanding that the same principles apply to antiparticles, ensuring that our conclusions are equally relevant to both.

The localized state for a particle, as described in eq. (\ref{eq:FWPOEGP}), naturally defines the momentum representation of a particle position operator in the FW representation: 
\begin{eqnarray}
\label{eq:FWPPO}
X^k_{FW} = i\partial_{p^k} - i\frac{p^k}{2E_{\textbf p}^2}.
\end{eqnarray} 
This definition is established by requiring that the position operator satisfies the eigenvalue equation:
\begin{eqnarray}
\label{eq:FWPOEG}
X^k_{FW} e^{i{ p}\cdot{ x}} u_{+\epsilon}(p,\lambda) &=& x^k e^{i{ p}\cdot{ x}} u_{+\epsilon}(p,\lambda).
\end{eqnarray}
The FW position operator $X^k_{FW}$ includes an additional term $-i p^k/(2E_{\textbf p}^2)$, unlike in non-relativistic QM, which is essential for making the FW position operator $X^k_{FW}$ Hermitian under the scalar product in Eq. (\ref{eq:NWFWP}) using the Lorentz invariant integral measure on the mass shell. Similarly, the particle position operator $X^k_P$ in the Dirac representation, as indicated in Eq. (\ref{eq:PPO}), also is not just $i\partial_{p^k}$ to guarantee its Hermitian nature. Notably, the form of the position operator in Eq. (\ref{eq:FWPPO}) aligns with that for the Klein-Grodon (scalar) particle case in NW \cite{NW49}.


The FW position operator $X^k_{FW}$ transforms to 
\begin{eqnarray}
\label{eq:FMDC}
U^\dagger_{FW}({\textbf p}) X^k_{FW} U_{FW}({\textbf p})
\end{eqnarray}
in the Dirac representation. As demonstrated by NW \cite{NW49}, the position operator derived from the locality condition in Eq. (\ref{eq:NWFWP}) should be unique. In fact, we have shown that the position operator in Eq. (\ref{eq:FMDC}), which is the FW mean position operator \cite{FW50}, is equivalent to $X^k_P$ in the Dirac representation \cite{Our21PO}. Consequently, $X^k_{FW}$ should be interpreted as the radius vector, embodying the operator equivalent of the coordinate vector $x^k$ within the FW framework, as described in FW \cite{FW50}, even though it has an additional term. The classical relation ${\textbf v}={\textbf p}/E_{\textbf p}$.which can be directly derived from the commutator $-i[H_{FW}, X^k_{FW}] $ for $H_{FW}=\beta E_{\textbf p}$, further supports this interpretation. 
Next we consider the Lorentz transformation properties of the position eigenstates in Eq. (\ref{eq:FWPOEG}) in the FW representation, simply called FW position eigenstates. 
When applying Lorentz transformations $\Lambda$ to the FW representation, the Lorentz transformation of the FW position eigenstate is described by:     
\begin{eqnarray}
\label{eq:LTFWPEGST}
e^{i{ p}\cdot{ x}} u_{+\epsilon}(p,\lambda) \longrightarrow S_{FW}(\Lambda) e^{i{\Lambda  p}\cdot{ x}} u_{+\epsilon}(p,\lambda).
\end{eqnarray}
Here, $S_{FW}(\Lambda)$, which facilitate the transformation of the FW spinor from the frame of momentum ${\textbf p}$ to the transformed frame of momentum $\Lambda {\textbf p}$, is defined as: 
\begin{eqnarray}
\label{eq:LTFW}
S_{FW}(\Lambda)=U_{FW}({\textbf P})M(\Lambda) U^\dagger_{FW}(\Lambda^{-1}{\textbf P}).
\end{eqnarray}
Using this $S_{FW}(\Lambda)$, the Lorentz transformed state of the FW position eigenstate in Eq. (\ref{eq:LTFWPEGST}) becomes:
\begin{eqnarray}
U_{FW}(\Lambda {\textbf p})M(\Lambda) U^\dagger_{FW}({\textbf p}) e^{i{\Lambda  p}\cdot{ x}} u_{+\epsilon}(p,\lambda) =  e^{i{\Lambda  p}\cdot{ x}} \sum_{\lambda'} \mathcal{D}_{\lambda' \lambda}(R) \,u_{+\epsilon}(\Lambda p,\lambda'),
\end{eqnarray}
where $u_{+\epsilon}(\Lambda p,\lambda')= U_{FW}(\Lambda {\textbf p}) \psi_{+\epsilon}(\Lambda {\textbf p},\lambda')$. As a result, the FW position eigenstate in the original frame of momentum ${\textbf p}$ transforms into a position eigenstate in the transformed frame of momentum $\Lambda {\textbf p}$ under the Lorentz transformation $\Lambda$, similar to the Dirac representation case. This transformation ensures that the covariance requirement of the particle position eigenstate under Lorentz transformations is satisfied in the FW representation as well.  

In non-relativistic QM, the interpretation of $\psi^\dagger({\bf x})\psi({\bf x})$ as probability density for locating a particle at a specific position ${\textbf x}$ at time $0$ per unit volume is straightforwardly supported by the action of the position operator $X^k$ on the state $\psi({\bf x})$, satisfying:
\begin{eqnarray}
X^k \psi({\bf x}) = x^k \psi({\bf x}).
\end{eqnarray} 
In contrast, relativistic QM introduces a more nuanced action for the position operator. Specifically, the action of the FW particle position operator on the FW spinor field $\psi_{FW}({\bf x})$ is given by: 
\begin{eqnarray}
\label{eq:APOFW}
X^k_{FW} \, \psi_{FW}({\bf x}) = x^k \psi_{FW}({\bf x}) +  \frac{1}{8\pi} \int d^3{\textbf y} \frac{e^{-m \lvert{\bf x}-{\bf y}\lvert}}{\lvert{\bf x}-{\bf y}\lvert}\frac{\partial \psi_{FW}({\bf y})}{\partial y^k},
\end{eqnarray}
drawing a parallel to the position operator and its eigenstates for the Klein-Gordon particle as discussed in NW \cite{NW49}. 

In the Dirac representation, we have demonstrated \cite{Our21PO} that the action of the particle position operator $X_P^k$ on the Dirac spinor field $\psi_D({\bf x})$ closely mirrors the behavior seen in the FW case:
\begin{eqnarray}
\label{eq:APO}
X_P^k \, \psi_D({\bf x}) \approx x^k \psi_D({\bf x}) +  \frac{1}{8\pi} \int d^3{\textbf y} \frac{e^{-m \lvert{\bf x}-{\bf y}\lvert}}{\lvert{\bf x}-{\bf y}\lvert}\frac{\partial \psi_D({\bf y})}{\partial y^k}.
\end{eqnarray}
This approximation is justified by the equality of the norm of the position eigenstate between the Dirac and the FW representations. The presence of the second term in both Eqs. (\ref{eq:APOFW}) and (\ref{eq:APO}), significant on the scale of a few Compton wavelengths, arises from the deviation of position eigenstates from Dirac delta functions in coordinate space within relativistic QM \cite{NW49,Our21PO}.  

Consequently, the densities $\psi^\dagger_D(x)\psi_D(x)$ and $\psi^\dagger_{FW}(x)\psi_{FW}(x)$ can be interpreted as the probability density for finding a particle at a specific position ${\textbf x}$ at time $x^0$ per unit volume, albeit with the spatial resolution limited to a few Compton wavelengths. This framework aligns with the one-particle limit where attempting a position measurement with precision beyond the Compton wavelength may induce pair creation. 

\section{Required properties of the probability density}
\label{sec:RPP}

In this section, we delve into the physical requirements that underpin a valid probability density within the framework of relativistic QM. We focus on two fundamental properties essential for ensuring the physical consistency of probability interpretations: the continuity equation, which guarantees the conservation of probability, and Lorentz invariance, which ensures that probability measurements are consistent across all inertial frames. 


\subsection{Continuity equation}

A probability density $\rho$ should satisfy the continuity equation, which is of the form
\begin{eqnarray}
\frac{\partial}{\partial t} \rho + \boldsymbol{\nabla}\cdot{\textbf j}=0,
\end{eqnarray}  
to ensure the conservation of total probability. Here, ${\textbf j}$ is the probability current density. This equation is crucial as it guarantees that the total probability across the entire space is conserved over time, despite the probability density's fluctuations at specific points or within certain regions. Additionally, it ensures the local conservation of probability, meaning that any local change in the probability density must be accompanied by a corresponding flux of the probability current ${\textbf j}$ across space.

The continuity equation should be consistent with the wave equation in its respective representation space. Notably, the Dirac equation in the Dirac representation
\begin{eqnarray}
i \frac{\partial}{\partial t} \psi_{D}(x) = - i \boldsymbol{\alpha}\cdot{\boldsymbol{\nabla}} \psi_D(x) 
\end{eqnarray}
 derives a continuity equation \cite{PQMDirac}:
\begin{eqnarray}
\frac{\partial}{\partial t} \psi^\dagger_D(x)\psi_D(x) + \boldsymbol{\nabla}\cdot \psi^\dagger_D(x) \boldsymbol{\alpha}\psi_D(x)=0,
\end{eqnarray} 
with the Dirac probability 4-current density being $(\psi^\dagger_D(x)\psi_D(x), \psi^\dagger_D(x) \boldsymbol{\alpha}\psi_D(x))$.  

Similarly, in the FW representation, the FW probability current density ${\textbf j}_{FW}$ emerges as a formal solution to the continuity equation \cite{Foldy56}:
\begin{eqnarray}
\label{eq:FWCE}
\frac{\partial}{\partial t} \left( \psi^\dagger_{FW}(x) \psi_{FW}(x)\right) = -\boldsymbol{\nabla}\cdot {\textbf j}_{FW}.
\end{eqnarray}
${\textbf j}_{FW}$ is non-local, as illustrated by the FW representation of the Dirac equation:
\begin{eqnarray}
\label{eq:FWHE}
i \frac{\partial}{\partial t} \psi_{FW}(x) = \gamma^0 \sqrt{-\boldsymbol{\nabla}^2+m^2} \, \psi_{FW}(x).
\end{eqnarray} 
This non-locality could make the interpretation of the localized particle for the FW spinor field $\psi_{FW}(x)$ awkward. 
However, the non-local characteristic of ${\textbf j}_{FW}$ itself does not undermine the integrity of interpreting $\psi^\dagger_{FW}(x)\psi_{FW}(x)$ as a valid probability density, as rigorously documented by Foldy \cite{Foldy56}.

Consequently, the probability 4-current densities in both the Dirac and FW representations satisfy the continuity equation, making them candidates for valid probability 4-currents. However, the non-locality in the FW probability 4-current density necessitates a deeper investigation into their Lorentz transformation properties. This exploration is crucial for discerning the comparative validity of the Dirac and FW probability densities.

\subsection{Lorentz invariant probability}

The interpretation of $\psi^\dagger_{FW}(x) \psi_{FW}(x)$ as a probability density meets fundamental challenges concerning the localization of a particle within a given volume. Consider a particle confined within a small box of volume $V$ defined as $d^3{\textbf x}$. Then the probability of finding a particle inside this volume is $\rho V$, hence the probability density $\rho$ must satisfy:  
\begin{eqnarray}
\label{eq:TPRB}
 \rho \, d^3 {\textbf x} =1
\end{eqnarray}
to ensure that the probability of finding the particle inside $d^3 {\textbf x}$ is one. To ensure the probability density $\rho$ is consistent across all inertial frames, it must transform as the time (0-th) component of a Lorentz $4$-vector. 
 
As is well-known, the Dirac probability density $\psi^\dagger_D(x) \psi_D(x)$ corresponds to the time component of the Lorentz 4-vector $\bar{\psi}_D(x)\gamma^\mu \psi_D(x)$, where the Dirac gamma matrices are defined as $\gamma^0=\beta$ and $\gamma^k=\gamma^0 \alpha^k$, and the Dirac adjoint field is $\bar{\psi}_D(x) = \psi^\dagger_D(x)\gamma^0$. The behavior of the Dirac probability density under a Lorentz transformation, $\Lambda$, is as follows \cite{PQMDirac}:    
\begin{eqnarray}
\label{eq:DPDLT}
\psi^\dagger_D(x) \psi_D(x) \longrightarrow &\,\,& \bar{\psi}_D(\Lambda^{-1}x) M^{-1}(\Lambda) \gamma^0 M(\Lambda) \psi_D(\Lambda^{-1}x)\\ \nonumber  
&=& \Lambda^0_{\phantom{\mu}\mu} \bar{\psi}_D(\Lambda^{-1}x)  \gamma^\mu \psi_D(\Lambda^{-1}x),
\end{eqnarray}
using the relation:
\begin{eqnarray}
\label{eq:GALT}
M^{-1}(\Lambda) \gamma^0 M(\Lambda)= \Lambda^0_{\phantom{\mu}\mu} \gamma^\mu.
\end{eqnarray}
This equation shows the covariant transformation of the Dirac probability density as the time component of a Lorentz 4-vector. 

We can check the consistent condition from Eq. (\ref{eq:DPDLT}), which states that the Dirac probability density defined in the transformed frame is the same as the $0$-th component of the transformed Dirac probability $4$-current density:
\begin{eqnarray}
\label{eq:CONSCON}
\psi^\dagger_D(\Lambda^{-1}x)\psi_D(\Lambda^{-1}x)=\bar{\psi}_D(\Lambda^{-1}x)\gamma^0 \psi_D(\Lambda^{-1}x),
\end{eqnarray}
because $\bar{\psi}_D(\Lambda^{-1}x)  \gamma^\mu \psi_D(\Lambda^{-1}x)$ is the Lorentz 4-vector defined by Eq. (\ref{eq:DPDLT}) in the transformed frame.

Similarly, to exhibit Lorentz covariance, the FW probability density should transform formally as:
\begin{eqnarray}
\label{eq:FWDLT}
\psi^\dagger_{FW}(x)\psi_{FW}(x) \longrightarrow \Lambda^0_{\phantom{\mu}\mu} \psi^\dagger_{FW}(\Lambda^{-1}x)J^\mu_{FW}\psi_{FW}(\Lambda^{-1}x),
\end{eqnarray}  
where $J^\mu_{FW}$ is introduced as a $4$-vector current operator, providing the FW probability $4$-current density $\psi^\dagger_{FW}(x)J^\mu_{FW}\psi_{FW}(x)$. To ensure that the FW probability density serves as the $0$-th component of the FW probability $4$-current density, it should satisfy the consistency condition that the FW probability density $\psi^\dagger_{FW}(\Lambda^{-1}x)\psi_{FW}(\Lambda^{-1}x)$ in the transformed frame should be equal to the $0$-th component of $\psi^\dagger_{FW}(\Lambda^{-1}x)J^\mu_{FW}\psi_{FW}(\Lambda^{-1}x)$:
\begin{eqnarray}
\label{eq:CONSFW}
\psi^\dagger_{FW}(\Lambda^{-1}x)\psi_{FW}(\Lambda^{-1}x) = \psi^\dagger_{FW}(\Lambda^{-1}x)J^0_{FW}\psi_{FW}(\Lambda^{-1}x).
\end{eqnarray}

The Lorentz transformation of the FW probability density is specified by the spinor representation $S_{FW}(\Lambda)$ of the Lorentz transformation $\Lambda$ in Eq. (\ref{eq:LTFW}) as: 
\begin{eqnarray}
\label{eq:LTFWPD}
\psi^\dagger_{FW}(x) \psi_{FW}(x) \longrightarrow {\psi}^\dagger_{FW}(\Lambda^{-1}x)  S_{FW}^{\dagger}(\Lambda)  S_{FW}(\Lambda) \psi_{FW}(\Lambda^{-1}x). 
\end{eqnarray} To compare with the know results of the Dirac case, it is convenient to express the detailed transformation of the FW spinor in terms of the Dirac spinor via $\psi_{FW}(x)=U^\dagger_{FW}({\textbf P})\psi_D(x)$. Through the action of the $S_{FW}(\Lambda)$, the transformation can be further unfolded as (Appendix \ref{sec:AppendixA}): 
\begin{eqnarray}
\label{eq:FWLT}
&&\psi^\dagger_{FW}(x) \psi_{FW}(x) \\ \nonumber &\phantom{\longrightarrow}& \phantom{\longrightarrow} \longrightarrow  \Lambda^0_{\phantom{\mu}\mu}\bar{\psi}_D(\Lambda^{-1}x)\gamma^\mu M^{-1}(\Lambda) \overleftarrow{U}^\dagger_{FW}({\textbf P})U_{FW}({\textbf P}) M(\Lambda)\psi_D(\Lambda^{-1}x),
\end{eqnarray}
highlighting the dependence on the Lorentz transformation matrix $\Lambda^0_{\phantom{\mu}\mu}$. This transformation shows the Wigner rotation given by the result of $M(\Lambda)\psi_D(\Lambda^{-1}x)$. Here $\overleftarrow{U}^\dagger_{FW}({\textbf P})$ acts to the left on the Dirac spinor $\bar{\psi}_D(\Lambda^{-1}x)$. Hence the transformed FW state is the Wigner-rotated state from the original spin state in the transformed frame.

For $\psi^\dagger_{FW}(x) \psi_{FW}(x)$ to act as the $0$-th component of the probability $4$-current density during the Lorentz transformation, the consistency condition in Eq. (\ref{eq:CONSFW}) requires:
\begin{eqnarray}
\label{eq:LTFWPD}
&&\psi^\dagger_{FW}(\Lambda^{-1}x) \psi_{FW}(\Lambda^{-1}x) \\ \nonumber &\phantom{\longrightarrow}& \phantom{\longrightarrow} = \bar{\psi}_D(\Lambda^{-1}x) \gamma^0 \overleftarrow{U}^\dagger_{FW}(\Lambda^{-1}{\textbf P}) U_{FW}(\Lambda^{-1}{\textbf P}) \psi_D(\Lambda^{-1}x).
\end{eqnarray}
This is similar to the consistency condition Eq. (\ref{eq:CONSCON}) in the Dirac representation. However, this equivalence is unattainable because the FW probability density defined in the transformed frame, $\psi^\dagger_{FW}(\Lambda^{-1}x) \psi_{FW}(\Lambda^{-1}x)$, does not incorporate the Wigner rotation resulting from $M(\Lambda)$ in the 0-th component among the transformed components of of the FW probability density on the right-hand side of equality in Eq. (\ref{eq:LTFWPD}). This reveals the absence of covariance in the transformed FW probability density. 

This discrepancy highlights a fundamental challenge: the inability to preserve a consistent probabilistic interpretation of a particle's location within a specific volume across different inertial frames. In other words, a particle confined to a box in one reference frame can be found outside that box in another reference frame. This result questions the validity of the FW probability density $\psi^\dagger_{FW}(x) \psi_{FW}(x)$ as a probability density in a relativistic setting.


\section{Conclusion}
\label{sec:Conc}

We explored the recently discussed question of which density, either the Dirac or the FW one, provides a consistent probabilistic interpretation in relativistic QM. We examined the concept of a localized state and its corresponding position operator, which is required to interpret a density as the probability of finding a particle per unit volume at a certain position ${\textbf x}$, similar to non-relativistic QM. 

We obtained localized states and their corresponding position operators by using the NW criteria in both the Dirac and FW representations. It was shown that the position operators in both representations make the density interpretable as the probability of finding a particle per unit volume within a few Compton wavelengths, which could be considered as the one-particle limit.

We also investigated the Lorentz covariance of the particle position in the two representations. We found that the localized eigenstates of the particle position operator in both representations are covariant under Lorentz transformations, which also guarantees the covariance of the particle position operators through the covariance of the position eigenvalues. This Lorentz covariance is a desirable property for a proper position operator in relativistic QM. 

However, the validity of the two densities as probability densities is determined by examining two commonly required properties for a probability density: the continuity equation and the invariance of the total probability of finding a particle within a small box. We found that the Dirac density satisfies both properties, whereas the FW density satisfies the continuity equation with a non-local probability current density, making the interpretation of a localized particle unclear. Additionally, we have shown that the FW probability density does not act as the $0$-th component of the probability $4$-current density, which does not guarantee the invariance of the total probability of finding a particle over a volume element across all reference frames. Consequently, the FW density fails to be a proper probability density.

Therefore, we conclude that the Dirac density is the proper probability density for representing the probability of finding a particle per unit volume at a specific position and time in relativistic QM. Our results provide a clear and consistent interpretation of the probability density for a massive spin 1/2 particle in relativistic QM, which will be beneficial for future applications.

While these observations underscore the invalidity in interpreting the FW probability density as a proper probability density in relativistic QM, they do not diminish the FW representation's utility or its fundamental equivalence to the Dirac representation via unitary transformation. The spectrum and other physical properties remain consistent across both representations due to their unitary equivalence. This analysis highlights the importance of selecting the appropriate representation for specific physical insights or calculations.




\acknowledgments
 This work was supported by a research grant from Seoul Women's University (2024-0068).

\appendix

\section{Derivation of Eq. (\ref{eq:FWLT})} \label{sec:AppendixA}

The Lorentz transformation of the FW spinor field $\psi_{FW}(x)$ under $\Lambda$ becomes
\begin{eqnarray}
S_{FW}(\Lambda)\psi_{FW}(\Lambda^{-1}x)= U_{FW}({\bf P})M(\Lambda) \psi_D(\Lambda^{-1}x) 
\end{eqnarray}
from Eqs. (\ref{eq:LTFWPEGST}) and (\ref{eq:LTFW}). Hence its adjoint becomes
\begin{eqnarray}
\psi^\dagger_D(\Lambda^{-1}x)M^\dagger(\Lambda) \overleftarrow{U}^\dagger_{FW}({\textbf P})=\psi^\dagger_D(\Lambda^{-1}x)\gamma^0 M^{-1}(\Lambda) \gamma^0 \overleftarrow{U}^\dagger_{FW}({\textbf P}),
\end{eqnarray}
where we use $\gamma^0 \gamma^0=1$ and $\gamma^0 M^\dagger(\Lambda)\gamma^0=M^{-1}(\Lambda)$. Using Eq. (\ref{eq:GALT}), it becomes 
\begin{eqnarray}
&&\psi^\dagger_D(\Lambda^{-1}x)\gamma^0 M^{-1}(\Lambda) \gamma^0 M(\Lambda)M^{-1}(\Lambda)\overleftarrow{U}^\dagger_{FW}({\textbf P}) \\ \nonumber 
&=&\psi^\dagger_D(\Lambda^{-1}x) \gamma^0 \Lambda^0_{\phantom{\mu}\mu}\gamma^\mu M^{-1}(\Lambda)\overleftarrow{U}^\dagger_{FW}({\textbf P}).
\end{eqnarray}
Now, it is easy shown that the FW probability density transforms as in Eq. (\ref{eq:FWLT}).


\end{document}